\documentclass[twocolumn,showpacs,superscriptaddress,floatfix,preprintnumbers,amsmath,amssymb,prb]{revtex4-1}

\usepackage{graphicx}
\usepackage{dcolumn}
\usepackage{bm}
\usepackage{undertilde}
\usepackage{verbatim}

\begin{document}

\title{Level crossings and zero-field splitting in the \{Cr$_8$\}-cubane spin-cluster by inelastic neutron scattering and magnetization studies}
\author{D. Vaknin}
\affiliation{Ames Laboratory, Department of Physics and Astronomy, Iowa State University, Ames, Iowa 50011, USA}
 \email{vaknin@ameslab.gov}
\author{V. O. Garlea}
\affiliation{Neutron Scattering Sciences Division, Oak Ridge National Laboratory, Oak Ridge, Tennessee 37831 USA}
\author{F. Demmel}
\affiliation{Rutherford Appleton Laboratory, ISIS Pulsed Neutron Facility, Chilton, Didcot, Oxon OX110QX, UK}
\author{E. Mamontov}
\affiliation{Neutron Scattering Sciences Division, Oak Ridge National Laboratory, Oak Ridge, Tennessee 37831 USA}
\author{H. Nojiri}
\affiliation{Institute for Materials Research, Tohoku University, Sendai 980-8577, Japan}
\author{C. Martin}
\affiliation{National High Magnetic Field Laboratory, Florida State University, Tallahassee, Florida 32310, USA}
\affiliation{Department of Physics, Florida State University, Tallahassee, Florida 32306}
\author{I. Chiorescu}
\affiliation{National High Magnetic Field Laboratory, Florida State University, Tallahassee, Florida 32310, USA}
\affiliation{Department of Physics, Florida State University, Tallahassee, Florida 32306}
\author{Y. Qiu}
\affiliation{NIST Center for Neutron Research, Gaithersburg, Maryland 20899, and University of Maryland, College Park, Maryland 20742, USA}
\author{P. K\"{o}gerler}
\affiliation{Ames Laboratory, Department of Physics and Astronomy, Iowa State University, Ames, Iowa 50011, USA}
\author{J. Fielden}
\affiliation{Ames Laboratory, Department of Physics and Astronomy, Iowa State University, Ames, Iowa 50011, USA}
\author{L. Engelhardt}
\affiliation{Department of Physics and Astronomy, Francis Marion University, Florence, South Carolina, 29501, USA}
\author{C. Rainey}
\affiliation{Department of Physics and Astronomy, Francis Marion University, Florence, South Carolina, 29501, USA}
\author{M. Luban}
\affiliation{Ames Laboratory, Department of Physics and Astronomy, Iowa State University, Ames, Iowa 50011, USA}

\date{\today}

\begin{abstract}
Inelastic neutron scattering in variable magnetic field and high-field magnetization measurements, at the milikelvin temperature range, were performed to gain insight into the low-energy magnetic excitation spectrum and the field-induced level crossings in the molecular spin cluster \{Cr$_8$\}-cubane.  These complementary techniques provide consistent estimates of the lowest level-crossing field. The overall features of the experimental data are explained using an isotropic Heisenberg model, based on three distinct exchange interactions linking the eight Cr$^{\text{III}}$ paramagnetic centers (spins $s = 3/2$), that is supplemented with a relatively large molecular magnetic anisotropy term for the lowest $S=1$ multiplet. It is noted that the existence of the anisotropy is clearly evident from the magnetic field dependence of the excitations in the INS measurements, while the magnetization measurements are not sensitive to its effects.
\end{abstract}

\maketitle

\section{Introduction}
Magnetic molecules are novel, physically realizable systems for exploring magnetic phenomena in low-dimensional magnetic materials\cite{Mannini2009,Schlegel2008,Bertaina2008,Koegerler2010,Gatteschi2006,Winpenny2004,Muller2001}.  Although these systems are very diverse, they possess the common characteristic of being achieved as crystalline samples of identical molecules, each containing a relatively small number of mutually interacting paramagnetic centers (``spins''). In most cases, intermolecular magnetic interactions are negligible, as compared to intramolecular exchange interactions, so that experiments on bulk samples primarily probe the properties of the common, individual molecular unit.   The small number of interacting spins in the basic molecular unit would suggest that it should be possible to formulate a theoretical model that can successfully rationalize experimental results.  However, as illustrated in the present work, in practice this task can present a significant challenge, especially when the experimental data are provided by complementary techniques.

In the present study we explore a \{Cr$_8$\}-cubane magnetic molecule, featuring eight interacting Cr$^\text{III}$ paramagnetic spin centers (spins $s = 3/2$)\cite{Atkinson1999,Fielden2010,Luban2003}. This system was also used to form a film on solid support by transfer of a monolayer at the air/water interface by the Langmuir-Blodgett technique\cite{Vaknin2001}. We report here detailed experimental results for several of the low-lying energy levels of this system as obtained by two complementary techniques, low temperature measurements of the magnetic field dependence of the magnetization and inelastic neutron scattering (INS). We find that the magnetization data can be accurately reproduced by a model based on purely isotropic Heisenberg exchange interactions between the eight Cr spins, but the field-dependent INS data are clearly inconsistent with a purely isotropic model.� Previous INS studies have demonstrated the effectiveness of the INS technique in directly determining the magnetic excitation spectrum of various magnetic molecules\cite{Hennion1997,Caciuffo1998,Zhong1999,Carretta2003,Waldmann2003,Garlea2006}. Hence we introduce anisotropy effects in a simplified manner, which is successful in explaining these measurements, while maintaining results that are consistent with the magnetization measurements.

In Sec.~\ref{Sec-Details}, various experimental details of the present work are described. These include a summary of the chemical synthesis of the \{Cr$_8$\}-cubane compound in Sec.~\ref{Sec-Synthesis}; and details of the experimental techniques are given for the magnetization and INS measurements in Secs.~\ref{Magnetization1} and~\ref{INS1}, respectively. Our experimental results are presented in Sec.~\ref{Sec-Results}, and they are analyzed in terms of two theoretical models in Sec.~\ref{Sec-Theory}. Specifically, the magnetization results of Sec.~\ref{Sec-Magnetization2} give rise to the isotropic Heisenberg model that is presented in Sec.~\ref{Sec-Isotropic}, and the INS results of Sec.~\ref{Sec-INS2} lead us to supplement that model as described in Sec.~\ref{Sec-Anisotropic}. Finally, in Sec.~\ref{Sec-Discussion} we summarize our present results and discuss open issues.

\begin{figure}[tbp]
\includegraphics[width=3.4in]{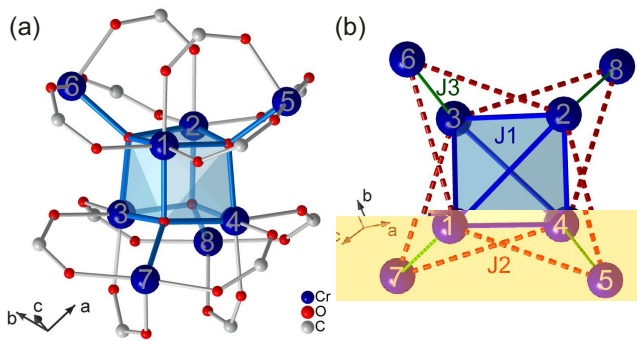}
\caption{\label{structure} (color online) (a) Structure of the \{Cr$_8$\}-cubane magnetic cluster. Phenyl groups and terminal benzoate ligands have been omitted for clarity. (b) scheme of the connectivity of the eight spin centers via superexchange pathways used in Sec.~\ref{Sec-Isotropic}.}
\end{figure}

\section{Experimental Details}\label{Sec-Details}

\subsection{Chemical Synthesis and Characterization}\label{Sec-Synthesis}

The \{Cr$_8$\}-cubane cluster [Cr$_8$O$_4$(O$_2$CPh)$_{16}$] and its deuterated analogue were synthesized following a modified protocol of a previously published procedure\cite{Atkinson1999} employing a crystallization method that uses fluorobenzene and acetonitrile as solvents in order to reliably obtain high-quality crystalline material\cite{Fielden2010}.  Samples were characterized by single-crystal X-ray diffraction and IR spectroscopy. We note that there are minimal structural differences between the present \{Cr$_8$\}-cubane compound [Cr$_8$O$_4$(O$_2$CPh)$_{16}$]$\cdot$4CH$_3$CN$\cdot$2H$_2$O (space group {\it C2/c}) and the previously published analogue\cite{Atkinson1999} [Cr$_8$O$_4$(O$_2$CPh)$_{16}$]$\cdot$2CH$_2$Cl$_2$ (space group {\it P2$_1$/c}) due to crystal packing effects.

\subsection{Magnetization Measurements}\label{Magnetization1}

Values of the magnetization \textit{M} versus external magnetic field \textit{H} for several low temperatures were obtained using two different measurement techniques, one based on a standard inductive method, and the second using a tunnel diode oscillator (TDO) method.
	Utilizing fast digitizers at the Tohoku University High Magnetic Field Laboratory, the inductive method provided data for $dM/dt$ and $dH/dt$ which were integrated to give results for \textit{M} versus \textit{H} for asymmetric half-cycle sweeps of duration 10 ms.  Samples were immersed in liquid $^4$He and $^3$He to maintain good thermal contact with the thermal bath.  The resulting data for \textit{M} versus \textit{H} as obtained for the up and down portions of the half cycle were in good agreement, indicating that hysteresis effects are negligible.  The measurements were performed on samples obtained from four different synthesis batches, and for several temperatures in the range 0.42 to 1.5 K.
	For the TDO measurements polycrystalline protonated samples were placed in the core of an inductive coil of a self-resonant LC tank circuit, powered by a tunnel diode.  The shift, $\Delta f(H)$, in the resonance frequency [11] of the coil due to the sample is proportional to the differential susceptibility $dM/dH$.  Measurements of $\Delta f$ versus $H$ were performed at fixed temperatures down to 300 mK using the $^3$He cooling system with 18 Tesla superconducting magnet at the National High Magnetic Field Laboratory (NHMFL).  The sweep rate was 0.3 T/min., and no hysteresis effects were observed between sweeping up and down the magnetic field.

\subsection{Neutron Scattering Measurements}\label{INS1}
Our INS measurements were performed both on deuterated and non-deuterated samples, with a base temperature of 60 mK.  The experiments reported here were performed on approximately 2.6 g of polycrystalline deuterated samples to reduce attenuation and incoherent scattering from hydrogen atoms. Samples were sealed in a copper holder under helium atmosphere.   The inelastic neutron spectra were collected using three different time-of-flight spectrometers: DCS at NCNR,\cite{DCS} OSIRIS at ISIS,\cite{OSIRIS} and BASIS at SNS, ORNL.\cite{BASIS} For the measurements at the DCS we used a fixed incident energy E$_i$ = 2.272 meV that yielded an energy resolution at zero energy transfer with a full width at half-maximum (FWHM) of approximately 60 $\mu$eV. Additional data were collected at the OSIRIS back-scattering spectrometer, with a fixed final neutron energy E$_f$ = 1.845 meV selected by a pyrolytic graphite PG(002) analyzer. The energy resolution that this instrument provided at zero energy transfer was approximately 20 $\mu$eV. The third set of high-resolution inelastic neutron measurements were performed using the BASIS spectrometer that uses backscattering neutron reflections from Si(111) analyzer crystals to select the final energy of the neutron of 2.08 meV. This instrument provides a $\sim4$ $\mu$eV energy resolution at the elastic position. The BASIS data treatment involved subtraction of a background spectrum measured separately with an empty Cu holder.  The INS results reported here are all from OSIRIS, as they are the more detailed ones, however the partial results from other instruments are consistent with the findings on OSIRIS.

\section{Experimental Results}\label{Sec-Results}

\subsection{Level crossings from magnetization}\label{Sec-Magnetization2}

\begin{figure}[ht!]
\includegraphics[width=3.0in]{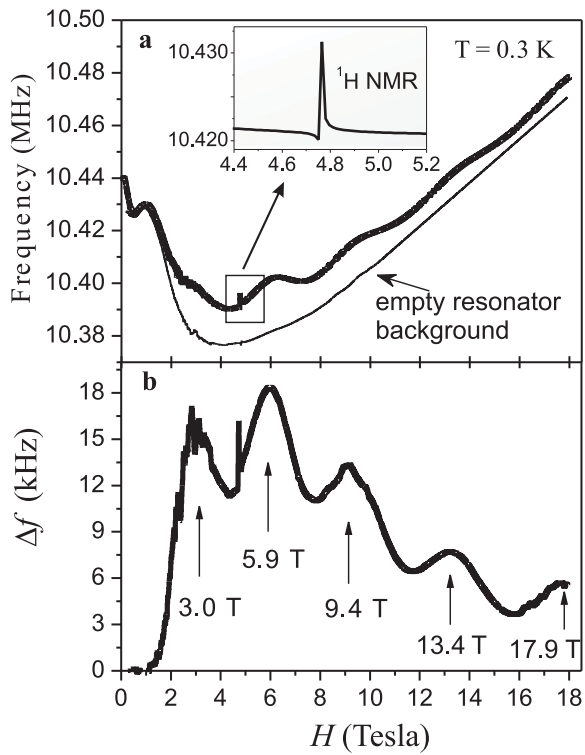}
\caption{\label{tdo} Variation with the magnetic field of the resonance frequency in a tunnel-diode oscillator.  (a) Raw data measured at $T=0.3K$. Inset shows the sharp line due to proton NMR at approximately 4.8 T. (b) The frequency shift of the intrinsic signal at T = 0.3 K after background subtraction.}
\end{figure}

Shown in Fig.\ \ref{tdo}(a) is the resonance frequency of the tunnel diode oscillator for magnetic fields up to 18 T as measured at 300 mK. The high sensitivity of the TDO method is demonstrated by the observation (inset of Fig.\ \ref{tdo}(a)) of the sharp peak at approximately 4.8 T that can be identified with the proton magnetic resonance absorption line.  Shown in Fig.\ \ref{tdo}(b) is the field dependence of the frequency shift $\Delta f$, equivalently $dM/dH$, due to the intrinsic {Cr$_8$}-cubane sample, as obtained by subtracting the background signal seen in Fig. 3(a) associated with an empty resonator.
	
Level crossing fields were identified by the appearance of peaks in the differential susceptibility $dM/dH$ versus $H$ as measured at fixed low temperatures, and these data are shown in Table\ \ref{Tab-Level-Xing}.  Specifically, columns 1-4 of the Table list the values of the level-crossing fields that were obtained using the inductive method. (These data were collected using four different samples, with fields in the range from 0 to 20 T.) The corresponding field values as obtained using the TDO method up to 18 T for a fifth sample are listed in column 5. The TDO peak at 17.9 T is only partially covered by the field sweep.    It should be emphasized that the two techniques provide similar results.

\begin{table}[!ht]
\caption{Comparison of the measured level crossings fields for five different samples and the prediction of the isotropic Heisenberg model (see Eq.\ (\ref{Eq-Hamiltonian})). Columns 1-4: pulsed-field method; column 5: TDO. The estimated error for all the pulsed-field entries is 0.15 T. For the TDO data the errors, listed in parenthesis, were obtained by a Lorentzian-based multi-peak fitting procedure.}
\begin{ruledtabular}
\begin{tabular} {cccccc}
 &&Pulsed-Field&&TDO& Heisenberg model \\
  &&(Tesla)&&& (Tesla) \\
\hline
2.6& 3.1& 2.8& 2.8 & 2.96(1) &2.9 \\
5.8& 6.1& 5.8& 6.3 & 5.88(1)&6.0 \\
9.4& 10.4& 9.1& 10.3 & 9.36(2)&9.3 \\
13.5& 14.6& 13.0& 14.1 & 13.38(3)&13.1 \\
17.9& 19.8& 17.8& 19.4& 17.9(1)&17.6 \\
\label{Tab-Level-Xing}
\end{tabular}
\end{ruledtabular}
\end{table}

\subsection{Inelastic neutron scattering}\label{Sec-INS2}

\begin{figure}[ht!]
\centering
\includegraphics[width = 0.4\textwidth]{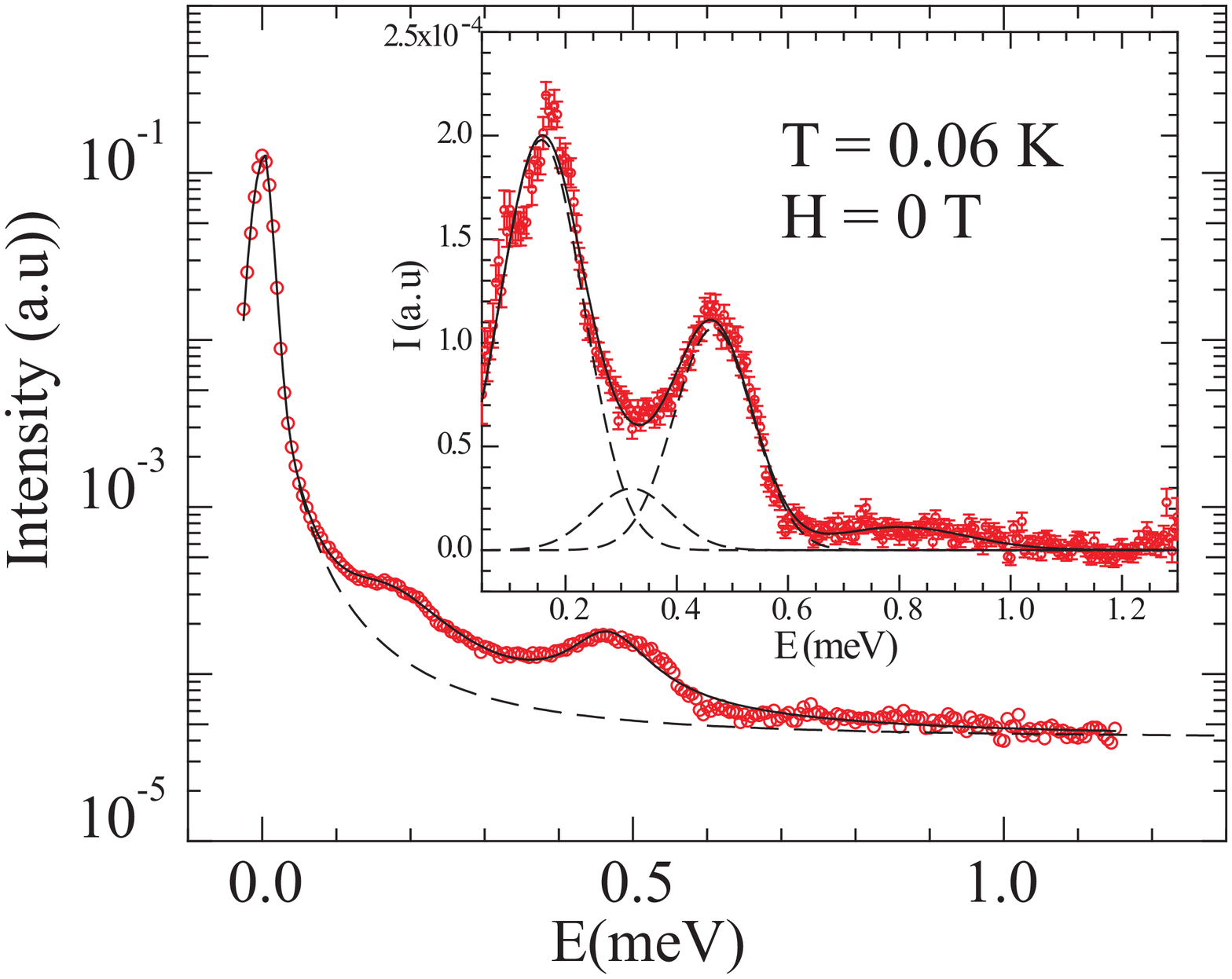}
\caption{\label{low-temp}(color online) (a) Inelastic neutron scattering spectrum at a nominal temperature of 60 mK. The intensity axis uses a logarithmic scale. The dashed line shows a best estimate of the nonmagnetic background (see text). Inset: The background subtracted scattering consists of three excitations at approximately 0.16, 0.32, 0.48 meV with similar widths, and a much broader and weaker excitation centered at $\sim$0.85 meV. (The error bars in all figures represent one standard deviation in the measurement.)}
\end{figure}

Figure~\ref{low-temp} shows intensities integrated over the $Q=0.3$ to 1.8 \AA$^{-1}$ range vs neutron energy loss, on a semilog plot at the lowest temperature achievable with the cryo-magnet, nominally $T=60$ mK. The large elastic peak has a FWHM of $\sim$0.021 meV consistent with the expected instrumental energy resolution for this configuration.\cite{Garlea2006}  Two broader peaks are clearly visible as superpositions on the otherwise bell-shaped resolution function.  To obtain the genuine magnetic spectra from the sample we reduce the data using a procedure we employed in a similar INS study of the \{Mo$_{72}$Fe$_{30}$\} magnetic molecule.\cite{Garlea2006}  The quasi-elastic term representing the instrumental resolution function, the incoherent and static disorder due to the sample (we name \textit{background curve}), is described by a sum of two co-centered peaks: a dominant Gaussian and a minor Lorentzian.  To obtain reliable parameters of the Gaussian/Lorentzian, the intensities at various temperatures are refined simultaneously, with extra peaks due to the magnetic spectra, while maintaining the same values of these parameters for the refinements of the various data sets at all measured temperatures.  The resulting \textit{background curve} (that includes non-magnetic contributions from the sample and holder) is shown as a dashed line in Fig.~\ref{low-temp}.  Subtracting the \textit{background curve} yields the genuine magnetic spectra as shown in the inset of Fig.~\ref{low-temp}.  The detailed analysis shows that the spectra at base temperature consists of three excitations at approximately 0.16, 0.32, 0.48 meV with similar widths, and a much broader and weaker excitation centered at $\sim$0.85 meV.

\begin{figure}[ht!]
\centering
\includegraphics[width = 0.3\textwidth]{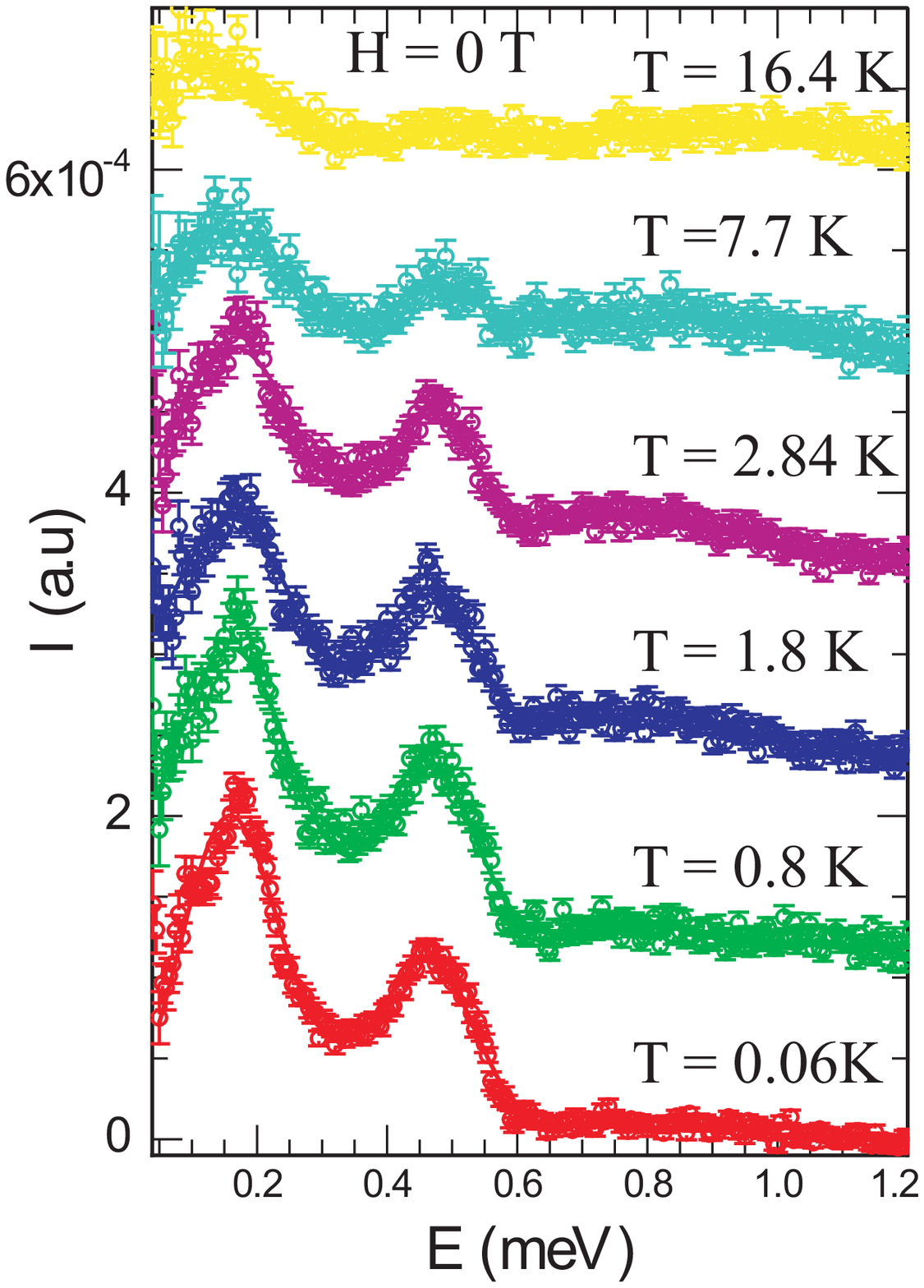}
\caption{\label{temp}(color online)  Magnetic excitations at various temperatures as indicated (shifted for clarity).  The data are obtained after background subtraction that includes the quasi-elastic non-magnetic contributions from the sample and the instrumental resolution, using a procedure similar to the one employed to obtain the inset of Fig.~\ref{low-temp}. }
\end{figure}
Figure~\ref{temp} displays the magnetic excitation spectra at various temperatures after the subtraction of the \textit{background curve} as described above.  The general effect of increasing temperature is in reducing the intensities of the excitations.  However, above $\approx 3$ K, the lowest excitation broadens and shifts to lower energies, whereas the other excitations broaden and almost disappear.  The broadening and shift to lower energy of the lowest excitation as the temperature is raised is commonly understood to be due to the occupation of excited states with access to many more allowed transitions some of which are associated with smaller gaps than the first excited states.  It should be noted that the experimental spectra at 60 and 800 mK are, within uncertainty, identical.  We therefore argue that a comparison of the INS measurements at 60 mK with the magnetization measurements at $T \sim 0.3 $ K is justified.

\begin{figure}[ht!]
\centering
\includegraphics[width = 0.46\textwidth]{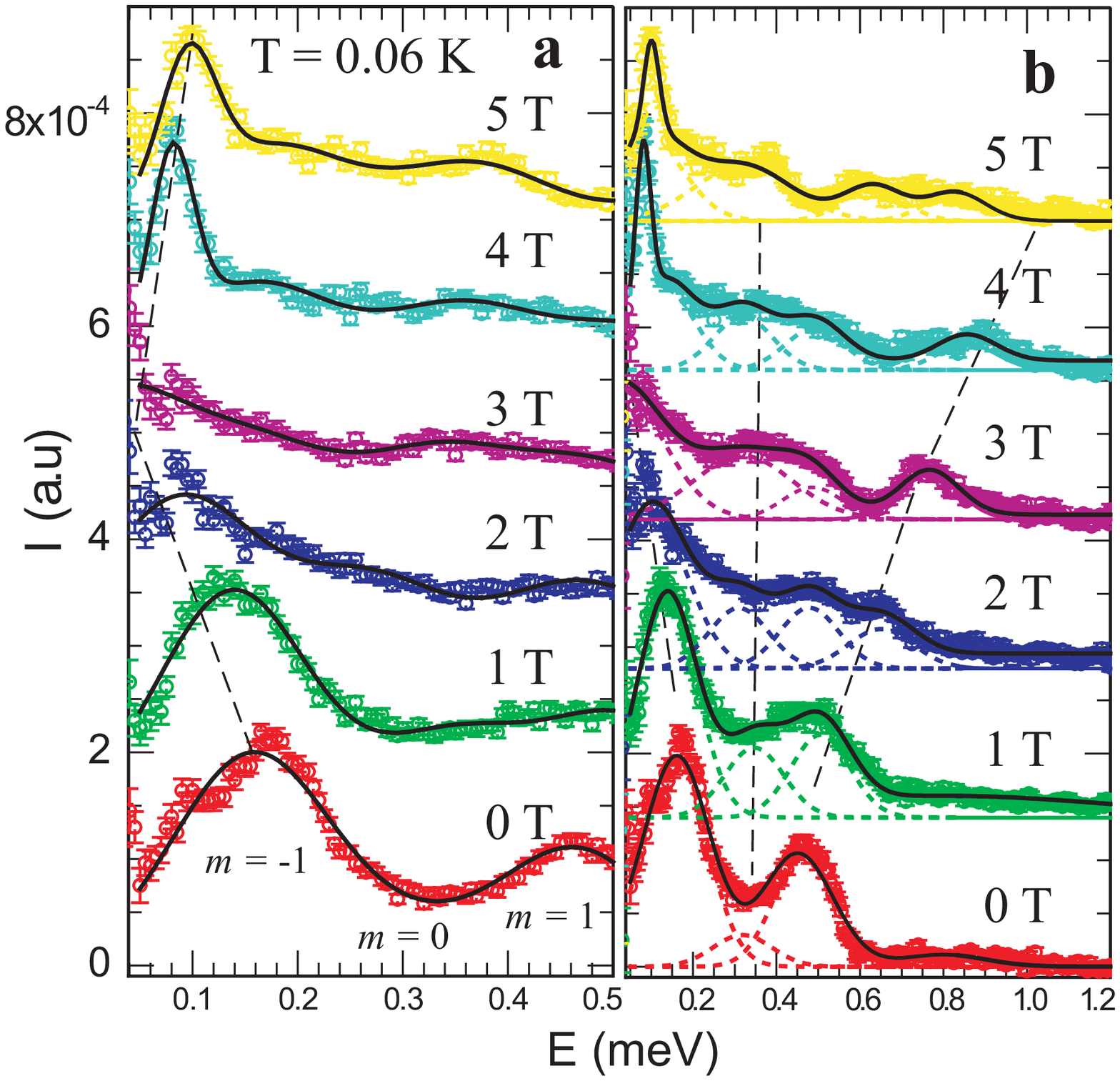}
\caption{\label{figH-ISIS}(color online) Magnetic spectra at $T = 0.06 $K at various applied magnetic fields obtained from INS measurements (shifted for clarity).  The spectra are obtained after the subtraction of the \textit{background} that includes the quasi-elastic non-magnetic contributions from the sample and the instrumental resolution, in a procedure similar to the one employed to obtain the inset of Fig.~\ref{low-temp}. The dashed lines show the progression of excitations with the increase of magnetic field.
}
\end{figure}

\begin{figure}[ht!]
\centering
\includegraphics[width = 0.42\textwidth]{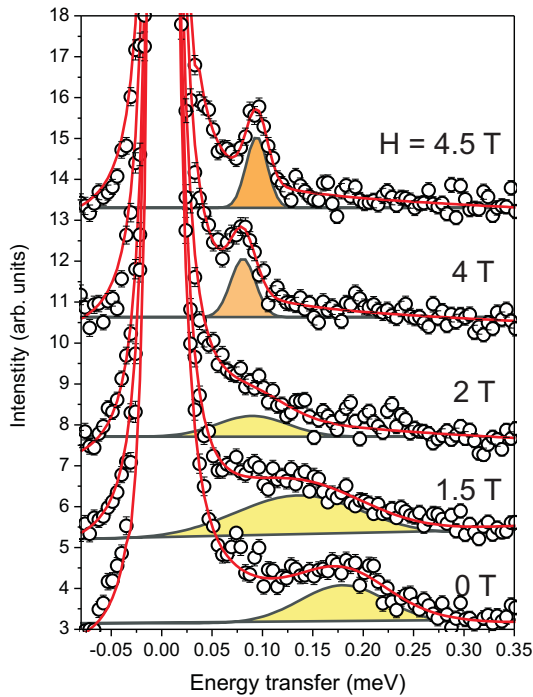}
\caption{\label{figH-Basis}(color online) INS raw data collected on BASIS (SNS) at various applied magnetic fields at $T = 1.5$ K. The high resolution setup (4 $\mu$eV) allows direct observation of the level crossing with increasing field, as the excitation shifts towards zero energy (below 2 Tesla) and the energy gap reopens at higher fields.}
\end{figure}
To confirm the magnetic origin of the excitations and to get more insightful characteristics of the spectra, we conducted the INS experiments under applied magnetic field at various temperatures. The \textit{background}-subtracted spectra at $T=0.06$ K are shown in Fig.~\ref{figH-ISIS} (a) and (b).  Similar results are obtained at $T=0.8$ K.  The lowest excitation (0.16 meV) is the most sensitive to the magnetic field.  This excitation shifts to lower energies and disappears between 2 and 3 T, and then reappears above $\sim 3$ T.  This can be interpreted as a direct observation of level crossing at approximately 2.8 T.  This is consistent with the first level crossing observed in the magnetization measurements ($\approx 2.8$ T).  In fact, INS raw data collected on BASIS (SNS) at various applied magnetic fields shows the same level crossing of the lowest excitation. The high resolution setup (4 $\mu$eV) allows one to clearly observe the level crossing, indicated by the reopening of the energy gap (Fig.\ \ref{figH-Basis}).

The behavior of the lowest excitations as a function of magnetic field is shown in Fig.~\ref{bands1}(b).   The excitation at 0.32 meV practically remains unchanged and broadens in the vicinity of the level crossing, and shifts to higher energies with the increase of magnetic field when the new ground state is established following the level crossing (traced as guide in Fig.~\ref{figH-ISIS}(b)).   The  0.48 meV  excitation shifts to higher energy  with increasing magnetic field and beyond the level crossing increases at a faster rate (shown in dashed line as guide in Fig.~\ref{figH-ISIS}(b)). At higher fields ($ \geq 4 $ T), excitations that are not accounted for by the $S=0$, and $S=1$ energy levels emerge. We hypothesize that they arise from excitations that involve the lowest $S=2$ multiplet.

\section{Comparison with Theory}\label{Sec-Theory}

\subsection{Isotropic Model}\label{Sec-Isotropic}

As a first attempt towards building a comprehensive theoretical model yielding results consistent with our experimental data, we utilized an isotropic Heisenberg Hamiltonian based on an initial assumption that the Cr$^{\text{III}}$ ions in their octahedral CrO$_6$ ligand field environments should be well approximated by pure spin-3/2 centers\cite{comment1}. Three distinct exchange constants are used to describe the interactions between the eight Cr$^{\text{III}}$ spin centers as shown in Fig. 1b. (Contrary to an earlier attempt\cite{Luban2003} in which pathways involving one $\mu$-oxo and one $\mu$-carboxylate group were distinguished from pathways involving one $\mu$-oxo and \emph{two} $\mu$-carboxylate groups, we here operate with a simplified model in which these exchange pathways are characterized by the same exchange constant.) The Hamiltonian, $\utilde{\mathcal{H}}$, of the spin system in the presence of an external magnetic field $H$ is given by
\begin{align}
\utilde{\mathcal{H}}=&J_1(\utilde{{\vec{s}}}_1\cdot\utilde{{\vec{s}}}_2
+\utilde{{\vec{s}}}_2\cdot\utilde{{\vec{s}}}_3
+\utilde{{\vec{s}}}_2\cdot\utilde{{\vec{s}}}_4
+\utilde{{\vec{s}}}_3\cdot\utilde{{\vec{s}}}_1 \nonumber \\
&+\utilde{{\vec{s}}}_3\cdot\utilde{{\vec{s}}}_4
+\utilde{{\vec{s}}}_4\cdot\utilde{{\vec{s}}}_1) \nonumber \\
&+J_2(\utilde{{\vec{s}}}_1\cdot\utilde{{\vec{s}}}_5
+\utilde{{\vec{s}}}_4\cdot\utilde{{\vec{s}}}_7
+\utilde{{\vec{s}}}_2\cdot\utilde{{\vec{s}}}_5
+\utilde{{\vec{s}}}_4\cdot\utilde{{\vec{s}}}_8
+\utilde{{\vec{s}}}_2\cdot\utilde{{\vec{s}}}_6\ \nonumber\\
&+\utilde{{\vec{s}}}_3\cdot\utilde{{\vec{s}}}_8
+\utilde{{\vec{s}}}_1\cdot\utilde{{\vec{s}}}_6
+\utilde{{\vec{s}}}_3\cdot\utilde{{\vec{s}}}_7) \nonumber \\
&+J_3(\utilde{{\vec{s}}}_1\cdot\utilde{{\vec{s}}}_7
+\utilde{{\vec{s}}}_4\cdot\utilde{{\vec{s}}}_5
+\utilde{{\vec{s}}}_2\cdot\utilde{{\vec{s}}}_8
+\utilde{{\vec{s}}}_3\cdot\utilde{{\vec{s}}}_6)\nonumber\\
&+g\mu_B\vec{H}\cdot\sum_{n=1}^8{\utilde{{\vec{s}}}_n},
\label{Eq-Hamiltonian}
\end{align}
where the individual spin operators are those for intrinsic spins $s = 3/2$, they are given in units of $\hbar$, and we used the value $g = 1.985$ for the spectroscopic splitting factor appropriate for Cr$^{\text{III}}$ spin centers. (Tildes, written below symbols, are used to denote quantum operators.) Inasmuch as the total spin operators $\utilde{S}^2$ and $\utilde{S}_z$ commute with $\utilde{\mathcal{H}}$, the eigenstates of these operators are described by quantum numbers $S$ and $M_S$ whose values range from 0 to 12 and from  $-S$ to $S$, respectively. Each multiplet has ($2S + 1$)-fold degeneracy when $H = 0$. In an external magnetic field the degeneracy is lifted due to a shift, $g\mu_{\tiny B}HM_S$, originating from the Zeeman term of Eq.~(\ref{Eq-Hamiltonian}). As the external field is increased from $H = 0$, the ground state changes successively from $S = 0$, $M_S = 0$  to $S = 1$, $M_S = -1$, etc., in integer steps of $S$ and $M_S$. Each of these changes of the ground-state quantum numbers is referred to as a ground state level crossing, and the field at which the ground state changes from $S - 1$ to $S$ will be denoted by $H_S$ . It is easy to show that the values of the $H_S$ can be determined using the relation $E_S-E_{S-1}=g\mu_BH_S$, $(S=1,2,3,...)$, where $E_S$ denotes the zero-field energy of the lowest-energy multiplet with quantum number $S$. (It is convenient to choose $E_0 = 0$.)

\begin{figure}[tbp]
\includegraphics[width=3.5in]{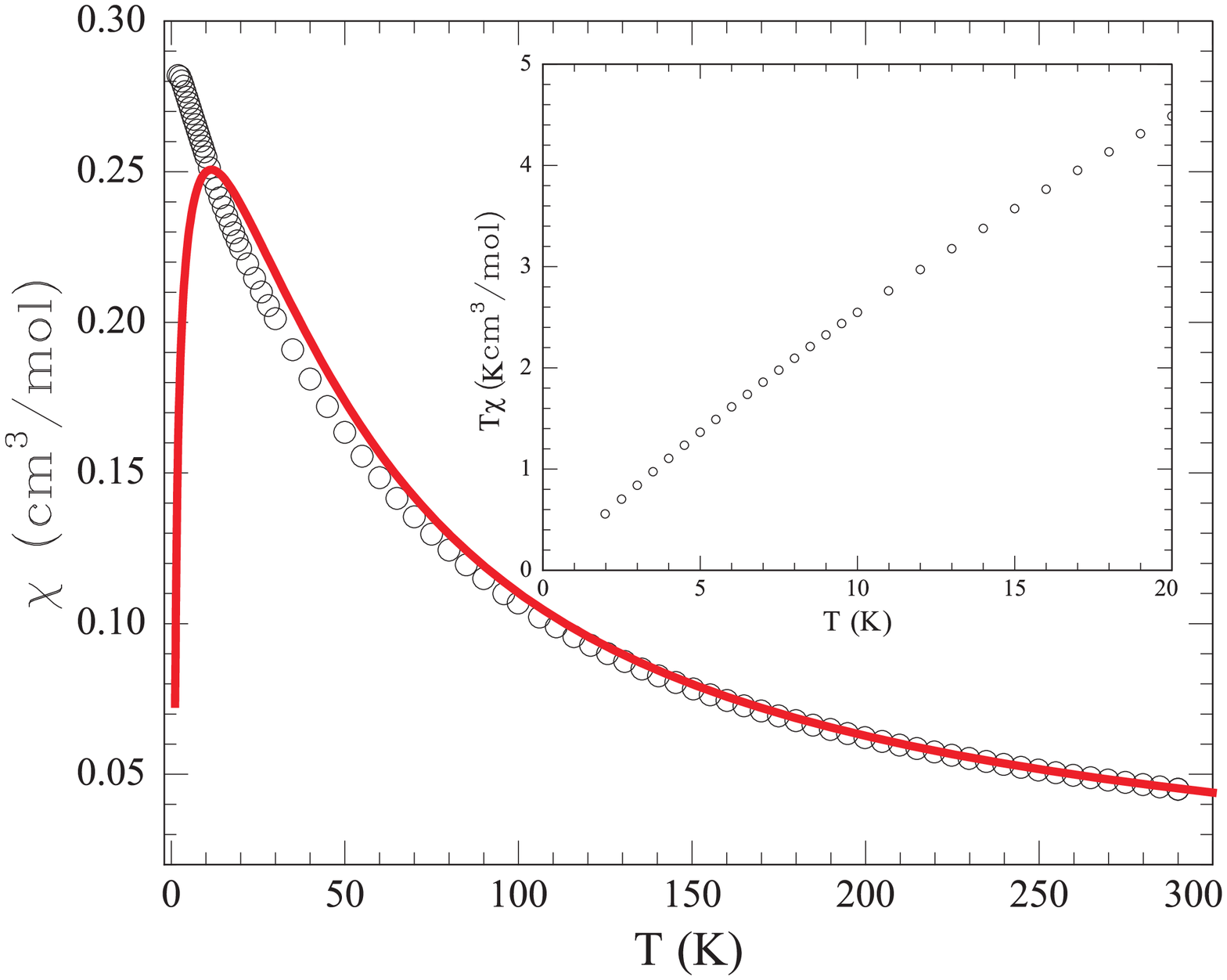}
\caption{\label{susce1} (color online) Molar magnetic susceptibility, $\chi=M/H$, of \{Cr$_8$\}-cubane for 0.5 T: Experiment (open circles) and theory (solid curve) as obtained using the isotropic Heisenberg model,  Eq.\ (\ref{Eq-Hamiltonian}) for optimized values of the exchange constants.   The approach of $T\chi \rightarrow 0$ as $T \rightarrow 0$ shown in the inset indicates that the ground state of the system is $S=0$ (see text for more detail). }
\end{figure}

To diagonalize the Hamiltonian of Eq.~(\ref{Eq-Hamiltonian}) we used the numerical package MAGPACK,\cite{MAGPACK} so as to provide the full set of energy levels as well as the temperature dependence of the weak-field susceptibility. The values of the exchange constants were first constrained using the method described in Ref.~\onlinecite{Luban2003} so as to reproduce the weak-field susceptibility (measured at 0.5 T using a Quantum Design MPMS magnetometer and shown in Fig.\ \ref{susce1}) above 100 K.  We then varied the values of the constrained exchange constants so as to optimize the fit between the calculated and measured values of the first five ground state level-crossing fields (see Table I). The resulting optimized values of the exchange constants are $J_1/k_B=32.1$ K, $J_2/k_B=2.2$ K, and $J_3/k_B=-23$ K.  The corresponding theoretical values of the ground state level-crossing fields are listed in the right-most column of Table~\ref{Tab-Level-Xing}, and it is noted that they are indeed in good agreement with our measured values

For the optimized isotropic Hamiltonian model the total angular momentum quantum number of the ground state for zero magnetic field is $S = 0$. This result is consistent with what can be inferred from the experimental weak-field susceptibility data. We recall the zero-field fluctuation formula that the limiting value of $T\chi$ for $T\rightarrow 0$ is given by $N_Ag^2\mu_B^2S(S+1)/(3k_B)$ for a ground state with quantum number $S$. (Here $N_A$ is Avogadro's number and $k_B$ is Boltzmann's constant.) Indeed, the data for $T\chi$, shown in the inset of Fig.\ \ref{susce1}, tend towards zero for $T \rightarrow 0$, whereas if $S = 1$ for the ground state the limiting low-temperature value would be $T\chi \approx 1$ cm$^3$ K/mol.

For completeness, we list here the values of the energy (in meV) and the $S$-quantum number for those levels of the isotropic model lying within 2 meV of the ground state: 0.338 ($S = 1$), 1.025 ($S = 2$), 1.173 ($S = 1$), 1.291 ($S = 0$), 1.660 ($S = 1$), 1.916 ($S = 2$) meV.

Finally, we remark that the susceptibility that is calculated using the optimized isotropic Hamiltonian (solid curve in Fig.\ \ref{susce1}) does not provide a good fit to the measured weak-field susceptibility at low temperatures. This fact and the field dependence of the INS data strongly suggest that non-Heisenberg terms play a significant role in the description of \{Cr$_8$\}-cubane.

\subsection{Including zero-field splitting}\label{Sec-Anisotropic}

Although the isotropic Hamiltonian of Eq.~(\ref{Eq-Hamiltonian}) is able to accurately reproduce the measured ground state level-crossing fields, our INS measurements provide clear evidence of additional non-Heisenberg interactions. Namely, as remarked above, the energy spectrum obtained from Eq.~(\ref{Eq-Hamiltonian}) will have a lowest $S=1$ multiplet that is three-fold degenerate (with energy $E_1$) for $H=0$, that splits into three levels ($E_1$,$E_1\pm g\mu_B H$) for $H \neq 0$. Such behavior would be detectable in an INS measurement as a single peak for $H=0$, that would split into three satellite peaks for $H>0$. However, as seen in Fig.~\ref{figH-ISIS}, our INS data show the presence of three (not one) distinct peaks in zero field that, for increasing field strengths, consist of 1) a low energy satellite, starting from 0.16 meV, that shifts to lower energy, 2) a high energy satellite, starting from 0.48 meV, that shifts to higher energy, and 3) a central peak at 0.32 meV.

We show now that the observed behavior of these three INS peaks in a magnetic field can be qualitatively understood in terms of an $S = 1$ triplet whose degeneracy in zero field has been lifted by an anisotropy mechanism rooted in the coordination environment of the four outer CrO$_6$ octahedra. Specifically, a closer inspection of the individual chromium ligand field environments in the molecular structure of {Cr$_8$}-cubane reveals that two types of coordination environments need to be differentiated\cite{Fielden2010}. Although the environment of the central four CrO$_6$ octahedra shows only relatively small deviations from the ideal O$_h$ symmetry (Cr-O bond distances only vary by 2\%, O-Cr-O bond angles by less than 10$^\circ$ from the 90$^\circ$ ideal), the outer CrO$_6$ groups experience a stronger distortion, primarily caused by terminal benzoate ligands, leading to very small (65$^\circ$) and very large (108$^\circ$) O-Cr-O bond angles. This significant distortion in turn results in single-ion zero-field splitting that we quantify using a phenomenological anisotropy model. Notably, the molecular symmetry clearly deviates from the tetrahedral ideal: The crystallographically imposed point group symmetry of the Cr$_8$-cubane molecules is $C_2$. The substructure of the eight Cr centers for example differs from a fully $T_d$-symmetric arrangement by more than 0.48 {\AA}. In this absence of tetrahedral or cubic symmetry, the single-ion anisotropy tensors of the Cr spin centers thus add up to a non-zero molecular anisotropy tensor, which supports the principal assumption of the molecular zero-field splitting.  We thus supplement the isotropic Hamiltonian of Eq.~(\ref{Eq-Hamiltonian}) with a general molecular anisotropy term of the form
\begin{equation}
\utilde{\mathcal{H}}^{\prime} = E_1\utilde{1}+ g\mu_B\vec{H}\cdot \utilde{{\vec{S}}}+d(\utilde{S}_z^2-\frac{2}{3}) + e(\utilde{S}_x^2-\utilde{S}_y^2),
\label{Eq-Second-Hamil}
\end{equation}
which we apply to the reduced (three-dimensional) subspace of the lowest $S=1$ triplet. The first term in Eq.~(\ref{Eq-Second-Hamil}) denotes the product of the field-free multiplet energy $E_1$ with the (3$\times$3) unit matrix, $\utilde{1}$; the Cartesian axes denote symmetry axes of a given molecule; and the three Cartesian components of the vector operator $\utilde{{\vec{S}}}$ are given by the standard (3$\times$3) matrix representation of angular momentum operators for the $S = 1$ manifold.  For zero magnetic field the eigenvalues of $\mathcal{H}^{\prime}$ are $E_1 -2d/3$, $E_1+d/3\pm e$.  We adopt the values $E_1 = 0.32$ meV, $d = 3e = 0.24$ meV thereby reproducing the energies (0.16 meV, 0.32 meV, and 0.48 meV) of the zero-field peaks that are observed in our INS measurements.
\begin{figure}[ht!]
\centering
\includegraphics[width = 0.4\textwidth]{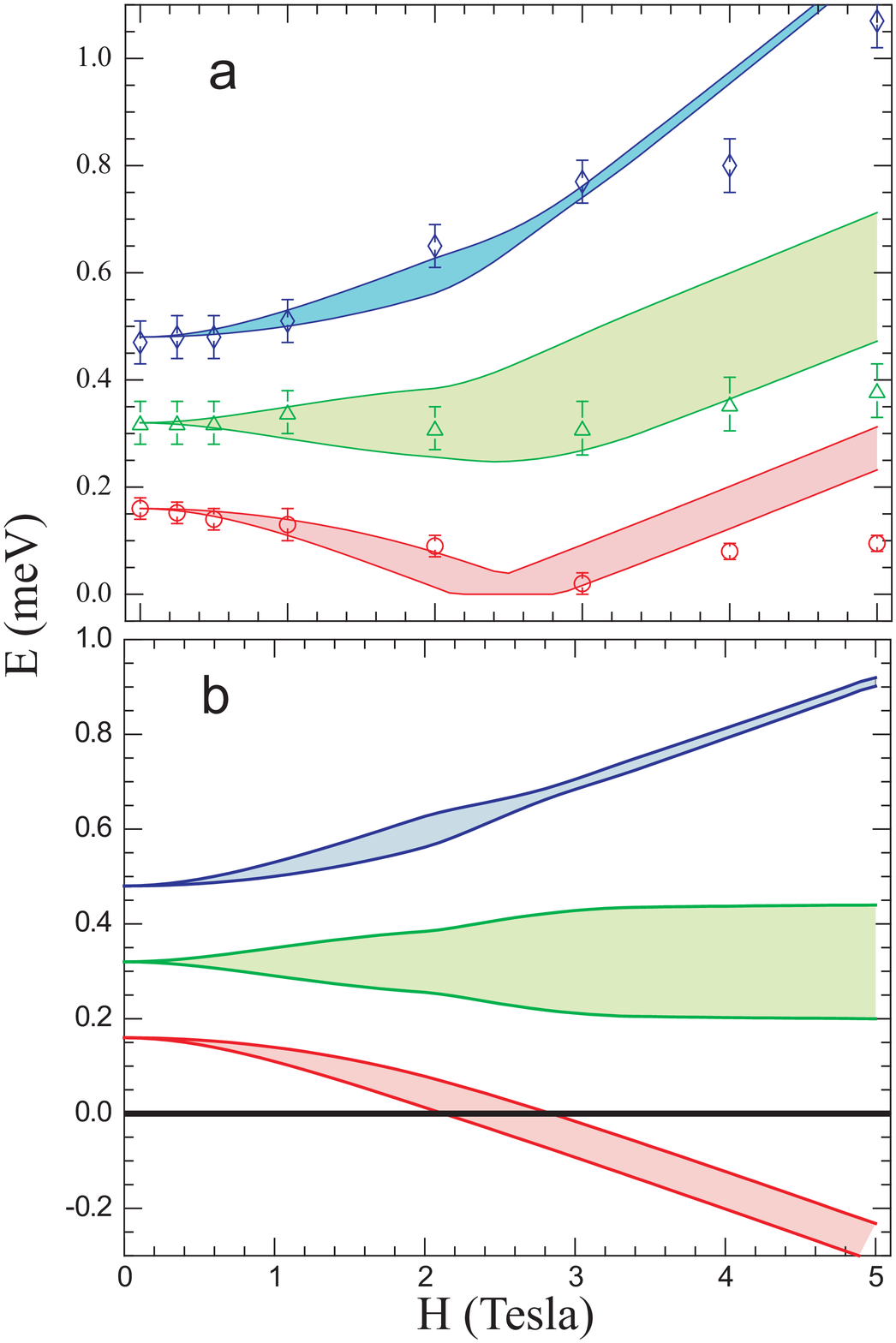}
\caption{\label{bands1}(color online) (a) Peak INS energies (symbols) and excitation bands (shaded) versus magnetic field. The excitation bands were obtained from Eq.~(\ref{Eq-Second-Hamil}) as described in the text. (b)  Energy bands versus magnetic field, obtained from Eq.~(\ref{Eq-Second-Hamil}) as described in the text.  }
\end{figure}

Although the three zero-field eigenvalues of $\utilde{\mathcal{H}}^{\prime}$ are independent of a molecule's orientation, this is not the case for $H>0$. Therefore, in order to describe an actual polycrystalline sample of \{Cr$_8$\}-cubane in the presence of a magnetic field, we have diagonalized $\utilde{\mathcal{H}}^{\prime}$ for many ($10^6$) uniformly distributed orientations of the Cartesian axes for each value of $H$. Each orientation produces a different set of three eigenvalues, and as the number of orientations becomes very large, these form three continuous distributions (bands) of energies. To show the variation of these distributions with field, we have calculated the mean value of each distribution plus(minus) its standard deviation, which define the energy bands that are shown in Fig.~\ref{bands1}(b). It can be seen that the lowest energy band intersects the $S=0$ ground state in the range 2.2 T $< H <$ 2.8 T; i.e., for some orientations the level-crossing field is as small as 2.2 T, and for other orientations it is as large as 2.8 T. This range of fields compares favorably with the 2.6 - 3.1 T range of observed values of the first level-crossing field, that are listed in Table I.  Since Eq.~(\ref{Eq-Second-Hamil}) does not produce a unique level-crossing field, this suggests that at very low temperatures peaks in $dM/dH$ should be broader than the peaks that are observed in a system without appreciable anisotropy, while at higher temperatures, this effect will be hidden within the thermal broadening. The width of the peaks shown in Fig.~\ref{tdo}(a) are consistent with the width as calculated using the isotropic model, Eq.~(\ref{Eq-Hamiltonian}), for $T=300$ mK; but we expect that the measured widths would reveal the effect of anisotropy if the experiment were repeated at a much lower temperature.

As described above, we have used Eq.\ (\ref{Eq-Second-Hamil}) to obtain three field-dependent energy eigenvalues for each of 10$^6$ different molecular orientations. Including the (field-independent) $S=0$ eigenvalue, this gives $4 \times 10^6$ eigenvalues for each value of magnetic field strength. In order to compare these results with the field-dependence of the INS measurements, we note that at 60 mK (the temperature used for the measurements shown in Fig.\ \ref{figH-ISIS}) only the ground state of a given molecule is expected to have significant thermal occupation, so transitions between excited states should be negligible. Therefore, we have calculated the energy differences between the ground state and the three excited states for each of the 10$^6$ orientations and for each value of magnetic field strength. (As described above, the ground state is $S=0$ for $H < 2$ T, it is $S=1 $ for $H > 3$ T, and the value of the level crossing field is different for different orientations.) Because these energy differences depend on the orientation of a molecule, they generate three excitation bands, and these are shown in Fig.\ \ref{bands1}(a), along with the field dependence of the INS peaks.   It can be seen that all three excitation bands exhibit the same qualitative behavior as observed for the INS peaks. Specifically, the lowest band decreases toward zero for $H<2$ T, then increases for $H>3$ T; the middle band is roughly independent of field for $H<3$ T, then slowly increases with field for $H>3$ T; and  the highest band has a positive slope for all $H>0$.  The poorer agreement visible above 4 T may be due to zero-field splitting effects of the lowest $S=2$ multiplet. These effects become significant for field strengths in the vicinity of the second level crossing field ($\approx 6 $ T).

Finally, we have also considered how the temperature dependent magnetic susceptibility is affected by the observed zero-field splitting. Our data for the weak-field (0.5 T) susceptibility (Fig. 7), extending down to 2 K, appear to be reaching a maximum (0.28 cm$^3$/mol) somewhere in the temperature interval (0, 2) K. By contrast the theoretical susceptibility as obtained using the optimized isotropic Heisenberg model has a maximum, 0.25 cm$^3$/mol, that occurs at the much high temperature $T = 11.6$ K. As remarked earlier, we take this as the second piece of evidence, besides our INS findings, for the basic inadequacy of a strictly isotropic description of the system.  To estimate how the anisotropy terms of Eq. (2) influence the low-temperature ($< 5$ K) susceptibility, we have calculated $M/H$ versus $T$ using Eq. (2) including only the four lowest energy levels that are discussed above (energies 0, 0.16, 0.32, 0.48 meV). The effect of anisotropy is very large: The resulting susceptibility has a maximum, 0.16 cm$^3$/mol, at 2.1 K.  While it is gratifying that the peak temperature is consistent with that of the available experimental data, the low value of the maximum is unsatisfactory. We have attempted to resolve this discrepancy by assuming the presence of independent paramagnetic impurities, i.e., by adding a Curie term, $C/T$, to the susceptibility of the four-level model. (Specifically, we chose $C = 0.25$ cm$^3$-K/mol so as to achieve agreement between the theoretical and experimental values (0.28 cm$^3$/mol) of the susceptibility for $T =2$ K.) This attempt was unsuccessful since the resulting theoretical value for $T=5$ K is 0.17 cm$^3$/mol, as compared to the experimental value 0.27 cm$^3$/mol.  The inadequacy of the four-level model may be due to not having included the other low-lying energy levels, in particular the anisotropy-perturbed levels of the lowest $S = 2 $ multiplet.  Unfortunately, at the present time the form and the strength of those anisotropy terms are unknown.

\section{Summary}\label{Sec-Discussion}

The present article reports an analysis of the magnetic field dependence of the energy spectrum for the \{Cr$_8$\}-cubane magnetic molecule using multiple experimental techniques. Ground state level crossings were observed using both tunnel diode oscillator (TDO) measurements and pulsed-field magnetization measurements, and excitation energies were measured directly using inelastic neutron scattering (INS) spectroscopy. In particular, peaks in $dM/dH$ were observed in both the TDO and pulsed-field measurements, which we identify as ground state level-crossing fields. These field values (given in Table I) are used in conjunction with high temperature susceptibility data to determine the exchange constants for the isotropic Heisenberg Hamiltonian of Eq.~(\ref{Eq-Hamiltonian}).

We note that the TDO and pulsed-field measurements provide no hint of the need for anything other than an isotropic Heisenberg model, while it is clear from our INS measurements that such a model is incomplete. Specifically, we observed three INS peaks, and based on their field dependence, we are able to identify the energy levels as belonging to the lowest $S=1$ multiplet. These peaks show that this lowest S = 1 multiplet is split, even in zero magnetic field, which is unexplainable in terms of the Heisenberg Hamiltonian.\cite{Tuna1}  The magnitude of the zero-field splitting (0.16 meV) is comparable to that observed for the antiferromagnetic ring [Cr$_8$F$_8$Piv$_{16}$].\cite{Slageren2002}

We attribute the observed field dependence of the INS data as being due to the effects of anisotropy not included in the initial (Heisenberg) Hamiltonian, and we hence supplement that Hamiltonian with an additional perturbing term, given in Eq.~(\ref{Eq-Second-Hamil}), that incorporates the zero-field splitting of the lowest $S=1$ multiplet. We find that by including this anisotropy term, we are able to achieve a semi-quantitative understanding of the INS data, while maintaining results that are consistent with the magnetization measurements.


\begin{acknowledgments}
We thank R.~E.~P.~Winpenny and J. Schnack for valuable discussions. The work at the Ames Laboratory was supported by the Office of Basic Energy Sciences, U.S. Department of Energy under Contract No.  DE-AC02-07CH11358.  The research at Oak Ridge National Laboratory's Spallation Neutron Source,  was sponsored by the Scientific User Facilities Division, Office of Basic Energy Sciences, U. S. Department of Energy.  The work at the NHMFL was supported by NSF cooperative agreement Grant No. DMR-0654118 and NSF Grant No. DMR-0645408. The work at the NCNR is supported in part by the National Science Foundation under Agreement No. DMR-0454672.  L.E. acknowledges support from the FMU Professional Development Committee. H.N. acknowledges support by Grant-in-Aid on Priority Areas ``High Field Spin Science in 100 T'' (Grant No. 451) from MEXT, Japan.
\end{acknowledgments}


\end{document}